\documentclass[aip,jcp,reprint,longbibliography]{revtex4-2}

\usepackage{graphicx}
\usepackage{bm}
\usepackage{physics}
\usepackage{mathtools}
\usepackage{amssymb}
\usepackage{amsthm}
\usepackage{mathrsfs}
\usepackage{tensor}
\usepackage{tikz}
\usepackage{xcolor}
\usepackage{hyperref}
\usepackage{booktabs}
\usepackage{multirow}
\usepackage{enumitem}

\usepackage{adjustbox}
\usetikzlibrary{
   arrows.meta,
   positioning,
   calc,
   fit,
   backgrounds,
   shapes.geometric
}

\usepackage{listings}
\begin{document}

\title{
SAKE: Spectral Autodiff Kernel Expansion for Liouvillian Response Transport
}

\author{Eric R. Bittner}
\email{ebittner@central.uh.edu}
\affiliation{Department of Physics, University of Houston, Houston, Texas 77204, USA}
\affiliation{Institut Courtois \& D\'epartement de physique, Universit\'e de Montr\'eal, 1375 Avenue Th\'er\`ese-Lavoie-Roux, Montréal, Qu\'ebec H2V~0B3, Canada}

\author{Carlos~Silva-Acu\~na}
\email{carlos.silva@umontreal.ca}
\affiliation{Institut Courtois \& D\'epartement de physique, Universit\'e de Montr\'eal, 1375 Avenue Th\'er\`ese-Lavoie-Roux, Montréal, Qu\'ebec H2V~0B3, Canada}

\author{Hao Li}
\affiliation{Institut Courtois \& D\'epartement de physique, Universit\'e de Montr\'eal, 1375 Avenue Th\'er\`ese-Lavoie-Roux, Montréal, Qu\'ebec H2V~0B3, Canada}

\author{Sim\'on~Paiva-Ortega}
\affiliation{Institut Courtois \& D\'epartement de physique, Universit\'e de Montr\'eal, 1375 Avenue Th\'er\`ese-Lavoie-Roux, Montréal, Qu\'ebec H2V~0B3, Canada}

\date{\today}

\begin{abstract}

We introduce the Spectral Autodiff Kernel Expansion (SAKE), a differentiable computational framework for transporting nonlinear spectroscopic response between neighboring quantum dynamical models. Rather than recomputing multidimensional spectra independently for each Hamiltonian or Liouvillian, SAKE constructs local transport expansions about a reference model by combining forward-mode automatic differentiation with Duhamel transport theory. Automatic differentiation generates first-, second-, and third-order derivatives of the parameter-dependent Liouvillian, which are assembled into a pathway transport operator that maps the nonlinear response of a reference model onto neighboring systems. The framework is validated for a four-level excitonic dimer possessing an $\mathfrak{su}(2)\times\mathfrak{su}(2)$ symmetry by comparing second- and third-order transported pathway operators with exact projected transport matrices obtained from direct calculations. The third-order expansion accurately reproduces the projected transport operator and its associated pathway mixing. Beyond providing an efficient computational strategy, the transport operator reveals how coherent and dissipative perturbations redistribute amplitude among double-sided Feynman pathways, exposing mechanistic information that is not directly apparent from the nonlinear spectrum. SAKE thereby establishes a differentiable computational framework for nonlinear spectroscopy that supports efficient local parameter exploration, sensitivity analysis, and future inverse-design applications.

\end{abstract}

\maketitle

\section{Introduction}
\label{sec:intro}
Multidimensional spectroscopies provide some of the most powerful
tools available for probing quantum dynamics in condensed-phase and
molecular systems. By correlating excitation, waiting, and detection
intervals, techniques such as two-dimensional electronic spectroscopy
(2DES), two-dimensional infrared spectroscopy (2DIR), and nonlinear
magnetic resonance experiments resolve electronic couplings, coherence
transfer pathways, population relaxation, and environmental
fluctuations that are inaccessible to linear response
measurements~\cite{Mukamel1995,Jonas2003,Cho2008,Fuller2015,
Brixner2004,Hamm2011}. Central to the interpretation of these
experiments is the Liouville-space pathway picture introduced and
systematized by Mukamel and co-workers
\cite{Mukamel1995,Mukamel2000}, in which the measured signal is
expressed as a coherent sum of double-sided Feynman pathways generated
by successive light--matter interactions and propagated by an
effective Liouvillian
\cite{Mukamel1995,Jonas2003,Brixner2005,Cho2008,Engel2007}.
Although this formalism has been extraordinarily successful, neighboring Hamiltonians or Liouvillians generally require complete recomputation of the nonlinear response. Even small parameter changes therefore demand repeated construction of the Liouville-space propagators and response pathways, making exploration of multidimensional parameter spaces computationally expensive.

Our recent work established the geometric and transport setting for this
problem. Quasistatic work in an open quantum system can be written as a
curvature flux over control space \cite{Bittner:2026aa}, while stationary
response separates into metric and antisymmetric curvature sectors
\cite{JCP2a}. We subsequently showed that a reference spectral calculation
can be carried to a neighboring model through a Duhamel expansion of the
Liouvillian propagator \cite{JCP3}. The state--generator geometry developed by
Bittner and Silva-Acu\~na~\cite{JCP2b} place these local expansions on the
manifold of admissible stationary models and supplies its Levi--Civita
connection and covariant transport law.

The practical implementation of this transport theory, however, relied
on explicit symbolic differentiation of the Liouvillian to construct
the required transport tensors~\cite{JCP3}. Although effective for one or two
control parameters, symbolic differentiation rapidly becomes
impractical as the dimension of the control manifold and the order of
the transport expansion increase. Finite-difference approximations
avoid symbolic algebra but introduce truncation errors that propagate
through the transport hierarchy while requiring a rapidly growing
number of Liouvillian evaluations.

These observations motivate the central objective of the present work.
Rather than developing another perturbative expansion, we seek a computational framework
capable of generating the first-, second-, and third-order Liouvillian
derivative tensors required for local Duhamel transport automatically.

Automatic differentiation provides a natural computational framework for this objective. By propagating the chain rule through a differentiable computational graph, automatic differentiation computes exact parameter derivatives of numerical programs without symbolic manipulation or finite-difference approximations \cite{Rumelhart1986,GriewankWalther2008,Baydin2018}. As a result, it has become an important tool in electronic structure theory, molecular simulation, quantum dynamics, and scientific computing, where differentiable implementations of Hartree--Fock theory, density functional theory, coupled-cluster theory, Lindblad dynamics, nonequilibrium Green's functions, and quantum optimal control routinely generate high-order derivative information directly from computational graphs
\cite{TamayoMendoza2018,Abbott2021,Kasim2022DQC,
Zhang2022PySCFAD,Tan2023PROFESSAD,Zhang2024LNOCC,
Kasim2021DDFT,Nagai2022,vonStrachwitz2026DataEfficientXC,
Craig2024DiffLindblad,Gautier2025Dynamiqs,
Tosca2026VariationalBosonic,Heinrich2026QTraj,
Wang2022QOCMemory,Zhouyin2023ADNEGF,Sun2025BRAD}.

Despite these advances, existing applications primarily employ automatic
differentiation to optimize or analyze a fixed computational model. The present
work instead transports observable response between neighboring physical
models. The implementation deliberately differentiates only the
parameter-dependent Liouvillian, rather than the complete nonlinear-response
program. The resulting tensors are inserted into explicit Duhamel expansions
of the resolvent and propagator, making the approximation order and every
pathway contribution directly inspectable. Thus, automatic differentiation
automates the construction of the local transport expansion rather than
differentiating the complete nonlinear-response calculation.

Finally, we demonstrate the SAKE framework for a general excitonic
exchange model possessing an
$\mathfrak{su}(2)\!\times\!\mathfrak{su}(2)$
hierarchy through third order in the inter-dimer coupling and compare the transported nonlinear response against direct numerical calculations. After a one-time construction of the reference pathway basis and the required Liouvillian derivative tensors, neighboring responses are generated from local transport expansions rather than repeated explicit Liouville-space calculations. The resulting transport operators accurately reproduce the benchmark pathway dynamics over the parameter range considered and enable efficient iterative recovery of unknown model parameters through successive local transport updates.

Equally important, the transport operators retain direct physical interpretability. Rather than serving merely as numerical propagators, they define linear transformations on the operational space of Liouville-space pathways. Each matrix element quantifies the transfer of amplitude between distinct double-sided Feynman pathways induced by changes in the Hamiltonian or dissipative dynamics. Consequently, successive orders of the transport expansion identify the pathways coupled by a perturbation, quantify the strength of that coupling, and reveal how coherent and dissipative interactions redistribute spectral weight throughout the nonlinear response. The transport hierarchy therefore provides considerably more than an efficient computational algorithm: it exposes the microscopic mechanisms by which perturbations reshape the nonlinear spectrum.

In this sense, SAKE serves simultaneously as a computational framework and as a diagnostic tool for multidimensional spectroscopy. Whereas conventional brute-force calculations produce only the nonlinear response of a perturbed system, Liouvillian transport reveals how that response arises through the mixing of the underlying operational pathway basis. The transport operators therefore provide both an efficient computational surrogate and a physically transparent description of nonlinear response, exposing the microscopic pathway couplings responsible for the observed spectral evolution.
The remainder of this paper develops the implemented SAKE framework.
Section~\ref{sec:II} summarizes the Liouvillian derivative and local
transport construction.
Section~\ref{sec:III} presents automatic differentiation, the Duhamel
transport hierarchy, and model specification. Section~\ref{sec:IV} presents
the numerical implementation and validation, including assembly of the
pathway transport matrix and iterative parameter recovery.
Finally, Section~\ref{sec:V} discusses the computational and mechanistic
implications of pathway-space transport and outlines extensions to larger
systems, inverse spectral design, and related open-quantum-system applications.

\section{Liouvillian Transport and the SAKE Construction}
\label{sec:II}

SAKE constructs the response of a parameter-dependent quantum dynamical
model from local derivatives of its Liouvillian rather than rebuilding every
response pathway at every point in parameter space. Let
\(\boldsymbol{\lambda}\) denote the model parameters and
\(\mathcal L(\boldsymbol{\lambda})\) the corresponding Liouvillian. A
reference model at \(\boldsymbol{\lambda}_0\) supplies the pathway basis and
the resolvent and propagator factors from which the response is assembled.

Forward-mode automatic differentiation evaluates the coordinate derivative
tensors
\[
 \mathcal L_{,\mu},\qquad
 \mathcal L_{,\mu\nu},\qquad
 \mathcal L_{,\mu\nu\kappa}
\]
at the reference point. Explicit Duhamel identities then propagate these
derivatives through the resolvent and waiting-time propagators. Projection
onto the reference pathway basis produces the derivatives of the local
transport matrix, giving the order-\(m\) approximation
\[
 T^{(m)}(\boldsymbol{\lambda})
 =
 \sum_{n=0}^{m}\frac{1}{n!}
 \Delta\lambda^{\mu_1}\cdots\Delta\lambda^{\mu_n}
 T_{,\mu_1\cdots\mu_n},
 \qquad m\leq 3,
\]
where
\(\Delta\boldsymbol{\lambda}=\boldsymbol{\lambda}-\boldsymbol{\lambda}_0\).
Diagonal matrix elements describe the local renormalization of reference
pathways, while off-diagonal elements describe perturbation-induced pathway
mixing.

This paper focuses on that computational construction: automatic generation
of Liouvillian derivative tensors, their Duhamel assembly, and their numerical
validation. Bittner and Silva-Acu\~na develop the metric and curvature sectors
of stationary response in Ref.~\cite{JCP2a} and the state--generator embedding,
induced metric, response two-form, complex structure, and covariant transport
law in Ref.~\cite{JCP2b}. These structures interpret the local expansion geometrically
but are not required to evaluate the coordinate-derivative hierarchy in the
present SAKE-DT release.

The following section gives the differentiable graph and the explicit
Duhamel hierarchy used by the implementation.


\section{Automatic Differentiation of Liouvillian Transport}
\label{sec:III}

The SAKE construction replaces repeated solution of neighboring dynamical
models with a local transport expansion about a reference model. The
computational problem addressed in this section is to construct the
Liouvillian derivative tensors entering that expansion efficiently. Rather
than obtaining these tensors symbolically or by finite differences, the
present implementation evaluates ordinary coordinate derivatives of the
parameter-dependent Liouvillian using forward-mode automatic differentiation.

Automatic differentiation (AD), originally developed in the context of
algorithmic differentiation and later popularized through
back-propagation in neural networks, evaluates exact derivatives of a
differentiable computational graph by systematic application of the
chain rule
\cite{Rumelhart1986,GriewankWalther2008,Baydin2018}. Unlike
finite-difference methods, AD introduces no truncation error, and
unlike symbolic differentiation, it scales efficiently to large
computational graphs with only modest computational overhead.

Although AD has recently become an important tool in electronic
structure theory, molecular simulation, and open quantum dynamics,
existing applications use it primarily to optimize or analyze a fixed
computational model. Here we address a different problem. 
We employ automatic differentiation to construct the Liouvillian derivative tensors entering the local Duhamel transport expansion, thereby enabling efficient approximation of neighboring nonlinear responses.
The key observation is that the parameter-to-Liouvillian map and the response
pathway assembled from it define a differentiable computational graph. This
observation forms the computational foundation of the Spectral Autodiff
Kernel Expansion (SAKE).

\subsection{Differentiable Graph Representation}

The model specification exposes the sequence of differentiable operations
through which the measured response is constructed. This sequence provides
the computational structure needed for automatic differentiation.

\medskip

\textbf{Definition (Computational Graph).}
A computational graph is a directed acyclic graph (DAG) whose nodes
represent elementary differentiable operations and whose directed
edges encode the dependencies between intermediate quantities. The
terminal node represents the quantity to be evaluated, while every
intermediate node stores the information required for subsequent
application of the chain rule.

\medskip

\textbf{Theorem I (Differentiable Graph Representation).}
\emph{Let the model parameters map smoothly to a Liouvillian, and suppose
that the propagators, interaction maps, preparation maps, and
measurement functional depend differentiably on the model parameters
$\boldsymbol{\lambda}$. For a fixed experimental protocol and a fixed
numerical representation of the dynamics, the observable response can
be represented as a directed computational graph whose nodes are
differentiable operations and whose terminal node is the measured
response.}

\medskip

\begin{proof}
The parameter-to-Liouvillian map is smooth by assumption. Every subsequent
object appearing in the construction of the response is
obtained through compositions of differentiable maps. Consequently,
the numerical evaluation defines a finite directed acyclic
computational graph whose nodes represent differentiable operations.
Repeated application of the chain rule therefore permits automatic
differentiation of the graph to whatever order is supported by the
chosen differentiation backend and available computational resources.
\end{proof}

\medskip

\textbf{Corollary I (Automatic Differentiability of the Response).}
\emph{Under the hypotheses of Theorem I, automatic differentiation may
be applied to the response graph to evaluate parameter derivatives of
arbitrary order, limited only by differentiability, numerical
regularity, and computational resources.}

In the present SAKE-DT implementation, nested forward-mode automatic
differentiation is employed to compute first-, second-, and third-order
Liouvillian derivative tensors, which provide the input to the Duhamel
transport expansion developed below.

Automatic differentiation therefore eliminates the need for separate
analytic derivations of each term in the transport hierarchy. Once the
response has been expressed as a differentiable computational graph,
the derivative pathways are generated automatically by graph traversal
and repeated application of the chain rule.
The explicit Duhamel identities specify how these coordinate derivatives
enter the transport hierarchy, while automatic differentiation provides an
efficient way to evaluate them.

It is important to emphasize that the computational graph is not
assembled manually. Once the user specifies the Hamiltonian,
Lindblad collapse operators, control parameters, preparation, and
measurement operators, the SAKE implementation constructs the
corresponding Liouvillian and response graph algorithmically from the model
specification. The user therefore defines only the physical model and
experimental protocol; the differentiable computational graph and the
dependencies required for automatic differentiation and Liouvillian
transport are generated automatically by the software. A representative
workflow illustrating this procedure is given in
Appendix~\ref{appendix:C}.

In practice, the response is constructed through the sequence
\[
\mathcal L(\boldsymbol{\lambda})
\longmapsto
\mathcal G(\omega;\boldsymbol{\lambda})
\longmapsto
\mathcal O(\boldsymbol{\lambda}),
\]
where the intermediate computation may involve Liouville-space propagators,
interaction superoperators,
transport operators, and pathway amplitudes appropriate to the
particular spectroscopic protocol.

Figure~\ref{fig:elementary_dag} summarizes the implemented SAKE-DT workflow
and the directed acyclic graph traversed during one transport step. The
computation begins with the parameter-dependent
Liouvillian $\mathcal L(\boldsymbol{\lambda})$, which defines the
frequency-domain Green's function
$\mathcal G(\omega;\boldsymbol{\lambda})$, which propagates the input
state to the measured observable through projection onto the
measurement operator. More complex nonlinear spectroscopic response
calculations are obtained by repeated composition of this elementary
computational motif.

\begin{figure*}[t]
\centering
\begin{adjustbox}{max width=\textwidth}
\begin{tikzpicture}[
    >=Latex,
    font=\sffamily,
    line cap=round,
    line join=round,
    node distance=9mm and 11mm,
    mainnode/.style={
        rounded corners=2.5mm,
        minimum height=1.35cm,
        minimum width=3.1cm,
        align=center,
        inner sep=5pt,
        line width=0.9pt
    },
    modelnode/.style={
        mainnode,
        draw=black!65,
        fill=black!3
    },
    adnode/.style={
        mainnode,
        draw=orange!80!black,
        fill=orange!8
    },
    derivnode/.style={
        mainnode,
        draw=blue!65!black,
        fill=blue!7
    },
    duhamelnode/.style={
        mainnode,
        draw=violet!65!black,
        fill=violet!7,
        minimum width=4.0cm
    },
    pathwaynode/.style={
        mainnode,
        draw=green!50!black,
        fill=green!7,
        minimum width=3.8cm
    },
    outputnode/.style={
        mainnode,
        draw=orange!80!black,
        fill=orange!8,
        double,
        double distance=0.7pt,
        minimum width=4.0cm
    },
    forward/.style={
        -{Latex[length=2.6mm]},
        line width=1.0pt,
        draw=black!75
    },
    stagelabel/.style={
        font=\scriptsize\sffamily,
        text=black!65,
        align=center
    },
    annotation/.style={
        font=\small\sffamily,
        text=black!70,
        align=center
    }
]


\node[modelnode] (controls)
{
    \textbf{Control coordinates}\\[1mm]
    $\boldsymbol{\lambda}
    \mapsto
    \mathbf p(\boldsymbol{\lambda})$
};

\node[modelnode, right=of controls] (L)
{
    \textbf{JAX-compatible Liouvillian}\\[1mm]
    $\mathcal L(\boldsymbol{\lambda})$
};

\node[adnode, right=of L] (AD)
{
    \textbf{Forward-mode AD}\\[1mm]
    \texttt{jax.jacfwd}
};

\node[derivnode, right=of AD] (derivs)
{
    \textbf{Liouvillian derivatives}\\[1mm]
    $\mathcal L_{,\mu},\quad
    \mathcal L_{,\mu\nu},\quad
    \mathcal L_{,\mu\nu\kappa}$
};

\draw[forward]
(controls) --
node[above,stagelabel]
{(a)}
(L);

\draw[forward]
(L) --
node[above,stagelabel]
{(b)}
(AD);

\draw[forward]
(AD) --
node[above,stagelabel]
{(c)}
(derivs);


\node[
    duhamelnode,
    below=16mm of derivs
] (duhamel)
{
    \textbf{Duhamel propagator expansion}\\[1mm]
    resolvent and waiting-time derivatives
};

\node[
    pathwaynode,
    left=of duhamel
] (pathderivs)
{
    \textbf{Pathway derivatives}\\[1mm]
    $\partial_{\mu_1}\cdots\partial_{\mu_n}
    |P_a\rangle,\qquad n\leq 3$
};

\node[
    pathwaynode,
    left=of pathderivs
] (projection)
{
    \textbf{Reference-basis projection}\\[1mm]
    $T_a{}^b{}_{,\mu_1\cdots\mu_n}$
};

\node[
    outputnode,
    left=of projection
] (transport)
{
    \textbf{Taylor transport matrix}\\[1mm]
    $T^{(m)}(\boldsymbol{\lambda}),
    \qquad m=1,2,3$
};

\draw[forward]
(derivs.south)
-- ++(0,-7mm)
-| (duhamel.north);

\draw[forward]
(duhamel) --
node[above,stagelabel]
{(d)}
(pathderivs);

\draw[forward]
(pathderivs) --
node[above,stagelabel]
{(e)}
(projection);

\draw[forward]
(projection) --
node[above,stagelabel]
{(f)}
(transport);


\node[
    outputnode,
    below=13mm of $(projection.south)!0.5!(transport.south)$
] (pathways)
{
    \textbf{Transported pathway basis}\\[1mm]
    $P(\boldsymbol{\lambda})
    \simeq
    T^{(m)}(\boldsymbol{\lambda})P^{(0)}$
};

\node[
    outputnode,
    right=of pathways
] (signal)
{
    \textbf{Nonlinear response}\\[1mm]
    $S_{\mathrm{tr}}(\boldsymbol{\lambda})$
};

\draw[forward]
(transport.south)
to[bend right=12]
(pathways.north west);

\draw[forward]
(pathways) --
node[above,stagelabel]
{(g)}
(signal);


\node[
    annotation,
    below=5mm of signal
]
{
    The present implementation differentiates only
    $\mathcal L(\boldsymbol{\lambda})$; preparation,
    dipole operators, and pathway definitions are held fixed.
};

\end{tikzpicture}
\end{adjustbox}
\caption{
Implemented SAKE-DT workflow for local Liouvillian transport.
Dimensionless control coordinates are first mapped to the physical
model parameters, from which the JAX-compatible Liouvillian
$\mathcal L(\boldsymbol{\lambda})$ is constructed.  The labeled transitions indicate the successive computational operations: (a) mapping the dimensionless control coordinates to the physical model parameters; (b) forward-mode automatic differentiation of the parameter-dependent Liouvillian; (c) construction of the first-, second-, and third-order Liouvillian derivative tensors; (d) Duhamel assembly of the resolvent and waiting-time propagator derivatives; (e) projection of the resulting pathway derivatives onto the reference operational basis; (f) truncation of the transport expansion to the desired Taylor order $(m=1,2,3)$; and (g) projection of the transported pathway basis onto the detection operator to reconstruct the nonlinear spectroscopic response.
}
\label{fig:elementary_dag}
\end{figure*}

The directed edges represent the dependencies among the intermediate quantities. During the forward evaluation, the graph propagates the computation from the parameter-dependent Liouvillian to the measured observable. Automatic differentiation is applied to the Liouvillian node to generate the hierarchy of coordinate derivative tensors, which are subsequently assembled into pathway derivatives through the Duhamel expansion. Because each operation in the graph is differentiable, the transport hierarchy is constructed automatically without symbolic differentiation or finite-difference approximations.

For nonlinear optical spectroscopy, the terminal operation is the
measurement functional
\[
\mathcal O(\boldsymbol{\lambda})
=
N\,\mathrm{Tr}
\left[
\mu\,
\mathsf T_\gamma(\boldsymbol{\lambda})
\rho_0
\right].
\]
Consequently, the observable response is represented as the terminal
node of a differentiable computational graph, and derivatives with
respect to the model parameters are obtained by repeated application
of the chain rule through the graph. The Liouvillian transport
hierarchy is therefore generated automatically from the computational
graph itself, rather than by separate symbolic derivations of each
response tensor.

\subsection{Duhamel Expansion and Transport Hierarchy}

Theorem I establishes that the response is represented by a
differentiable computational graph, allowing automatic differentiation
to generate derivatives of the Liouvillian with respect to the control
parameters. The remaining task is to convert these local derivatives
into finite transport on the manifold of physical models.
This connection is provided by the Duhamel expansion. In our previous
work we showed that derivatives of the Liouvillian propagators may be
written as ordered insertions of Liouvillian derivative tensors within
the propagator itself. For the frequency-domain Green's function,
\begin{align}
    \mathcal G(\omega)=-\left[\mathcal L+(i\omega-\eta)I\right]^{-1},
\end{align}
the first derivative is
\[
\partial_\mu\mathcal G
=
-
\mathcal G
\,
(\partial_\mu\mathcal L)
\,
\mathcal G,
\]
with higher-order derivatives obtained by all ordered insertions of
$\partial_\mu\mathcal L$,
$\partial_{\mu\nu}\mathcal L$,
and higher derivative tensors. Analogous expressions follow for the
time-domain propagator through the Duhamel integral representation.

Automatic differentiation provides these Liouvillian derivatives
directly from the computational graph, while the Duhamel expansion
assembles them into the transport hierarchy. Consequently, every order
of the local transport operator is generated automatically without
requiring a separate symbolic derivation.

\subsection{Model Specification}

The SAKE framework begins from a parameter-dependent open quantum
system specified by a Hamiltonian
$H(\boldsymbol{\lambda})$, a set of Lindblad collapse operators
$\{C_k(\boldsymbol{\lambda})\}$, an initial state $\rho_0$, and the
interaction operator defining the experimental observable, typically
the transition dipole operator $\mu$. These quantities determine the
Liouvillian superoperator
\begin{align}
\mathcal L(\boldsymbol{\lambda})\rho
&=
-i[H(\boldsymbol{\lambda}),\rho]
\nonumber \\
&+
\sum_k
\left[
C_k(\boldsymbol{\lambda})\rho
C_k^\dagger(\boldsymbol{\lambda})
-
\frac12
\left\{
C_k^\dagger(\boldsymbol{\lambda})
C_k(\boldsymbol{\lambda}),
\rho
\right\}
\right],
\end{align}
which serves as the fundamental object of the transport
construction.

Internally, density operators are vectorized so that
$\mathcal L(\boldsymbol{\lambda})$ is represented as a matrix acting in
Liouville space. The implementation then treats the Liouvillian as a
differentiable map
\[
\boldsymbol{\lambda}
\longmapsto
\mathcal L(\boldsymbol{\lambda}),
\]
which forms the root node of the computational graph introduced in the
preceding section.

Automatic differentiation evaluates the required Liouvillian
derivative tensors directly from this graph,
\begin{align}
\mathcal L^{(0)}
&=
\mathcal L(\mathbf0),\\
\mathcal L_{,\mu}
&=
\left.
\frac{\partial\mathcal L}
{\partial\lambda^\mu}
\right|_{\mathbf0},\\
\mathcal L_{,\mu\nu}
&=
\left.
\frac{\partial^2\mathcal L}
{\partial\lambda^\mu\partial\lambda^\nu}
\right|_{\mathbf0},\\
\mathcal L_{,\mu\nu\kappa}
&=
\left.
\frac{\partial^3\mathcal L}
{\partial\lambda^\mu
\partial\lambda^\nu
\partial\lambda^\kappa}
\right|_{\mathbf0},
\end{align}
and higher derivatives as required. These tensors constitute the
elementary building blocks of the Duhamel transport hierarchy and are
generated automatically without symbolic differentiation or
finite-difference approximations.

The present SAKE-DT implementation employs ordinary coordinate derivatives of
the parameter-dependent Liouvillian together with explicit Duhamel expansions
to construct the local transport hierarchy. The response geometry of
Ref.~\cite{JCP2a} and the state--generator connection developed by Bittner and
Silva-Acu\~na~\cite{JCP2b} supply the geometric and covariant interpretation of
this expansion; no metric,
connection, or covariant derivative is evaluated by the current
implementation.

\section{Validation of the approach: pathway transport in the 
\texorpdfstring{$\mathfrak{su}(2)\times\mathfrak{su}(2)$}{su(2) x su(2)} dimer}
\label{sec:IV}

To validate the SAKE transport framework we consider the minimal
four-level excitonic dimer shown in
Fig.~\ref{fig:exact_transport_matrix}(a). 
The model consists of the site
basis
\[
|00\rangle,\quad
|10\rangle,\quad
|01\rangle,\quad
|11\rangle,
\]
in which the one-exciton states are coupled through both coherent
exciton exchange and incoherent population transfer. The coherent
contribution is generated by the Hamiltonian term
\[
H_J
=
J\left(
|10\rangle\langle 01|
+
|01\rangle\langle 10|
\right),
\]
while the dissipative Liouvillian introduces incoherent transfer
between the same states with rate $\kappa$. The pair
\[
\boldsymbol{\lambda}=(J,\kappa)
\]
therefore defines a minimal two-dimensional control manifold that
simultaneously probes coherent and dissipative modifications of the
nonlinear response.

\subsection{Benchmark validation}

The objective of the present calculations is to validate the computational implementation of SAKE rather than to investigate the spectroscopy of a particular physical model. The excitonic dimer provides a minimal two-parameter Liouvillian for which exact benchmark calculations, finite-difference derivatives, and autodifferentiated Duhamel expansions may all be compared directly. The benchmark therefore tests the accuracy of the implemented transport algorithm under controlled conditions.

Although employed here as a minimal benchmark, the four-level
$\mathfrak{su}(2)\times\mathfrak{su}(2)$ excitonic dimer captures the essential competition
between coherent excitonic coupling and dissipative population transfer
that underlies a broad class of excitonic systems studied by nonlinear
optical spectroscopy, including molecular H- and J-aggregates,
photosynthetic light-harvesting and reaction-center complexes,
conjugated polymers, donor--acceptor systems, and semiconductor
nanostructures
\cite{Kasha1963,Kasha1965,Mukamel1995,Cho2008,
Spano2010,SpanoSilva2014,HestandSpano2018,Blankenship2014}.

The calculations were carried out using the
\texttt{SAKE-DT} development branch of the
\texttt{Duhamel\_Transport} package, available through the repository
listed in the Data Availability Statement. The implementation follows
the modular workflow introduced in the previous section: specification
of the physical model, automatic differentiation of the Liouvillian,
construction of the Duhamel transport expansion, and reconstruction of
the nonlinear spectra from the transported pathway basis. A concise
description of the software architecture together with representative
pseudocode is provided in Appendix~\ref{appendix:C}.

The present benchmark evaluates the implementation exactly as used in the software. Forward-mode automatic differentiation generates Liouvillian derivative tensors through third order, which are assembled into Duhamel propagator derivatives and projected onto the reference pathway basis. No explicit metric, connection, or covariant differentiation is required in the current implementation.

The reference model is chosen as the uncoupled dimer,
\[
\boldsymbol{\lambda}=(J,\kappa)=(0,0),
\]
while the benchmark target is selected as a representative random point
within the local control manifold. No optimization or tuning of this
point was performed; rather, it serves as a generic neighboring model
whose parameters are assumed to be unknown. This reflects the
practical spectroscopic setting in which the measured response is
available, but the underlying Hamiltonian and dissipative parameters
must be inferred. The exact transport operator for this target then
provides the benchmark against which the SAKE expansion is compared.

\begin{table}[t]
\centering
\caption{Operational pathway basis used for the third-order transport
operator. The Liouville-space interaction strings denote the ordered
sequence of left ($L^\pm$) and right ($R^\pm$) actions of the dipole
superoperators on the density operator. Each string corresponds to a
single irreducible double-sided Feynman diagram.}
\label{tab:pathways}
\begin{tabular}{lll}
\hline
Pathway ket & Liouville interaction string & Physical process\\
\hline
$|P_1\rangle$ & RP: $L^-L^+R^+$ & (GSB)\\
$|P_2\rangle$ & RP: $L^-R^+L^+$ & (SE)\\
$|P_3\rangle$ & RP: $L^-R^+R^+$ & (ESA)\\
\hline
$|P_4\rangle$ & NRP: $R^+L^-L^+$ & (GSB)\\
$|P_5\rangle$ & NRP: $R^+L^-R^+$ & (SE)\\
$|P_6\rangle$ & NRP: $R^+R^-R^+$ & (ESA)\\
\hline
\end{tabular}
\end{table}

The pathway basis introduced here differs fundamentally from a
conventional basis of Liouville space. Each ket
\[
|P_n\rangle
\]
represents the contribution associated with a single double-sided
Feynman diagram of the nonlinear response. Operationally, it encodes
the ordered sequence of light--matter interactions,
Liouville-space propagators, and final detection defining one
irreducible nonlinear optical pathway. The pathway index therefore
labels Liouville-space processes rather than basis states of the
underlying Hilbert or Liouville spaces. Collectively, these pathway
kets form an operational basis for the nonlinear response.

For example, in a third-order experiment a pathway ket may correspond to an individual rephasing or nonrephasing double-sided Feynman diagram describing a ground-state bleach, stimulated-emission, or excited-state absorption process.
For the present four-level excitonic dimer, the operational basis consists of the six irreducible third-order double-sided Feynman diagrams satisfying the rephasing (RP) and nonrephasing (NRP) phase-matching conditions. These correspond to the ground-state bleach (GSB), stimulated-emission (SE), and excited-state absorption (ESA) pathways in each phase-matching sector.

The operational pathway basis used throughout the remainder of this work is summarized in Table~\ref{tab:pathways}.
Here $L^\pm$ and $R^\pm$ denote the action of the positive- and negative-frequency components of the dipole superoperator on the left and right sides of the density operator, respectively. The ordered sequence of these left and right actions uniquely specifies an irreducible Liouville-space response pathway and therefore defines one basis vector $|P_n\rangle$ of the operational pathway basis on which the transport operator acts.

For the reference model, the complete set of pathways is evaluated
independently and assembled into the reference pathway matrix
\[
P^{(0)}
=
\left[
|P_1(0)\rangle,\ldots,|P_N(0)\rangle
\right].
\]
The corresponding pathway basis is then constructed directly for a
target model,
\[
P(\boldsymbol{\lambda})
=
\left[
|P_1(\boldsymbol{\lambda})\rangle,\ldots,
|P_N(\boldsymbol{\lambda})\rangle
\right].
\]
The exact pathway transport operator is defined by
\[
P(\boldsymbol{\lambda})
=
T_{\mathrm{exact}}(\boldsymbol{\lambda})
P^{(0)},
\]
and is obtained by projecting the target pathways onto the reference
basis,
\[
T_{\mathrm{exact}}(\boldsymbol{\lambda})
=
P(\boldsymbol{\lambda})
\left[P^{(0)}\right]^+,
\]
where $(\cdot)^+$ denotes the Moore--Penrose pseudoinverse. For a
square, nonsingular pathway basis this reduces to the ordinary matrix
inverse. The resulting transport operator provides the benchmark
against which the SAKE expansion is compared.

\begin{figure*}
    \centering
    \includegraphics[width=\linewidth]{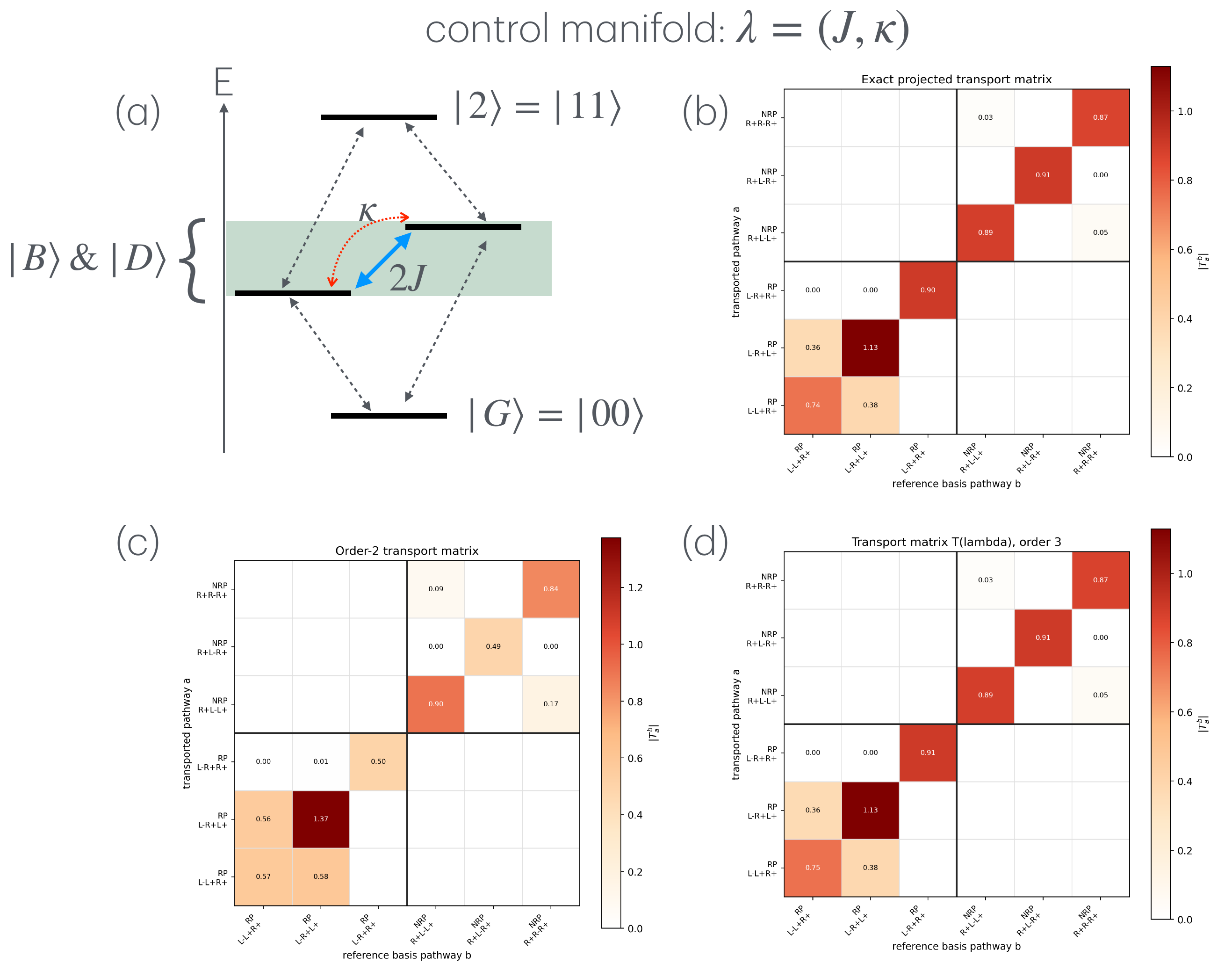}
   \caption{
\textbf{Validation of the Spectral Autodiff Kernel Expansion (SAKE) for the
$\mathfrak{su}(2)\times\mathfrak{su}(2)$ excitonic dimer. }(a) Energy-level diagram of the
model system. The control manifold is parameterized by the coherent
excitonic coupling $J$, which mixes the site-basis states
$|10\rangle$ and $|01\rangle$, and the incoherent population-transfer
rate $\kappa$, which couples the same states through the dissipative
Liouvillian. The ground and double-excited states are denoted
$|G\rangle=|00\rangle$ and $|2\rangle=|11\rangle$, while the
intermediate exciton states $|B\rangle$ and $|D\rangle$ are the bright
and dark superpositions of $|10\rangle$ and $|01\rangle$,
respectively.
(b)\textbf{Targeted projected pathway transport matrix}
$\left|T_a{}^{b}\right|$ obtained by directly evaluating the response
at a representative target point $(J,\kappa)$ and projecting the
resulting pathways onto the reference pathway basis.
(c,d) \textbf{Reconstructed} second- and third-order SAKE approximations
constructed from the Duhamel transport expansion using
autodifferentiated Liouvillian derivatives.
The columns label the double-sided Feynman pathways of the reference
system, while the rows label the transported pathways of the target
system. Heavy lines separate the rephasing (RP) and nonrephasing (NRP)
phase-matching sectors, which are transported independently. Diagonal
and near-diagonal matrix elements describe pathway survival and
renormalization, whereas off-diagonal elements quantify pathway mixing
induced by the coherent and dissipative interactions. The close
agreement between the exact transport matrix and the third-order SAKE
approximation demonstrates that the local Duhamel transport expansion
accurately reproduces both the dominant pathway amplitudes and their
mixing, thereby validating the transport operator itself rather than
only the resulting nonlinear spectrum.
}
    \label{fig:exact_transport_matrix}
\end{figure*}

Figure~\ref{fig:exact_transport_matrix} shows the magnitudes
$\left|T_a{}^b\right|$ for a representative target point selected in
the $(J,\kappa)$ control plane. The horizontal index labels the
reference pathway $b$, while the vertical index labels the transported
pathway $a$. The matrix is evaluated independently within the
rephasing and nonrephasing phase-matching sectors, producing the
block structure separated by the heavy lines.

The transport is not purely diagonal. Several pathways remain
dominated by a single reference contribution, with diagonal or
near-diagonal weights of approximately $0.87$--$0.91$. These elements
describe the continuous renormalization of pathways that retain their
identity as the coherent and dissipative couplings are introduced.
Other pathways exhibit substantial off-diagonal weight. Most notably,
the lower rephasing block contains coefficients of magnitude
$1.13$, $0.74$, $0.38$, and $0.36$, indicating strong mixing between
the corresponding reference pathways. The perturbed response therefore
cannot be described solely by rescaling the uncoupled pathways:
coherent exchange and incoherent transfer rotate the response within
the operational pathway space.

The exact transport matrix also preserves the phase-matching
decomposition. Rephasing pathways mix only with other rephasing
pathways, while nonrephasing pathways remain within the nonrephasing
sector. This block separation is a direct consequence of wave-vector
selection and provides an important structural constraint on the
transport construction.

The matrix elements provide a more sensitive diagnostic of the
Liouvillian deformation than the total spectrum alone. Diagonal
elements quantify the survival and renormalization of the original
double-sided Feynman diagrams, whereas off-diagonal elements identify
which optical processes are coupled by the added coherent and
dissipative interactions. In the present model, $J$ mixes amplitudes
through coherent exchange between $|01\rangle$ and $|10\rangle$,
while $\kappa$ transfers population incoherently between the same
one-exciton states. Their combined action therefore produces a
nontrivial redistribution of pathway weight even when the gross
spectral lineshape changes only modestly.



\subsection{Iterative parameter recovery with local SAKE charts}
\label{sec:iterative_inverse_benchmark}

The local character of the SAKE expansion suggests a natural strategy for
inverse parameter estimation.  Rather than requiring a single expansion about
the uncoupled reference to remain accurate over the entire control manifold,
we construct a sequence of overlapping local charts.  Within each chart the
third-order SAKE transport expansion provides an inexpensive surrogate for the
pathway states, while a direct resolvent calculation is used only to validate
the proposed update and, when accepted, to define the center of the next
chart.  The procedure therefore combines local differential transport with the
error control of a trust-region method.

For the asymmetric dimer we write
\begin{equation}
  \omega_1=\omega_0+\frac{\Delta\omega}{2},
  \qquad
  \omega_2=\omega_0-\frac{\Delta\omega}{2},
\end{equation}
and introduce the dimensionless controls
\begin{equation}
  \lambda_J=\frac{J}{\omega_0},
  \qquad
  \lambda_\kappa=\frac{\kappa}{\omega_0}.
  \label{eq:inverse_scaled_controls}
\end{equation}
The calculations reported here use $\omega_0=2.0$~eV and
$\Delta\omega=0.18$~eV, corresponding to site energies of 2.09 and
1.91~eV.  Rephasing and nonrephasing pathway-state vectors used in parameter
recovery were evaluated on the same $20\times20$ frequency grid at every
accepted chart center.  Spectra used for visualization were subsequently
evaluated on a $64\times64$ grid; thus, the inverse-recovery discretization does
not limit the resolution of the displayed line shapes.

Let $\mathbf P_{\rm tar}$ denote pathway states obtained by direct inversion at
a synthetic target and let $\mathbf P_n^{\rm SAKE}(\delta\boldsymbol\lambda)$
denote their third-order reconstruction in a chart centered at
$\boldsymbol\lambda_n$.  The local trial step is obtained from
\begin{equation}
 \delta\boldsymbol\lambda_n
 =\underset{\|\delta\boldsymbol\lambda\|_2\le R_n}{\arg\min}
 \frac{\left\|
 \mathbf P_n^{\rm SAKE}(\delta\boldsymbol\lambda)
 -\mathbf P_{\rm tar}\right\|_F^2}
 {\left\|\mathbf P_{\rm tar}\right\|_F^2},
 \label{eq:sake_inverse_objective}
\end{equation}
where $R_n$ is the current trust radius.  A direct calculation at the trial
point gives the acceptance ratio
\begin{equation}
 \rho_n=
 \frac{\chi^2_{\rm dir}(\boldsymbol\lambda_n)
       -\chi^2_{\rm dir}(\boldsymbol\lambda_n+\delta\boldsymbol\lambda_n)}
      {\chi^2_{\rm SAKE}(\boldsymbol\lambda_n)
       -\chi^2_{\rm SAKE}(\boldsymbol\lambda_n+\delta\boldsymbol\lambda_n)}.
 \label{eq:sake_trust_ratio}
\end{equation}
Steps that reduce the direct residual and satisfy $\rho_n\ge0.1$ are
accepted; the trust radius is then enlarged or reduced according to the
agreement between the predicted and direct decreases.  Positivity of
$\kappa$ was enforced together with the illustrative upper bound
$\kappa/\omega_0\le0.015$.

\begin{figure*}[t]
  \centering
  \includegraphics[width=0.96\textwidth]
  {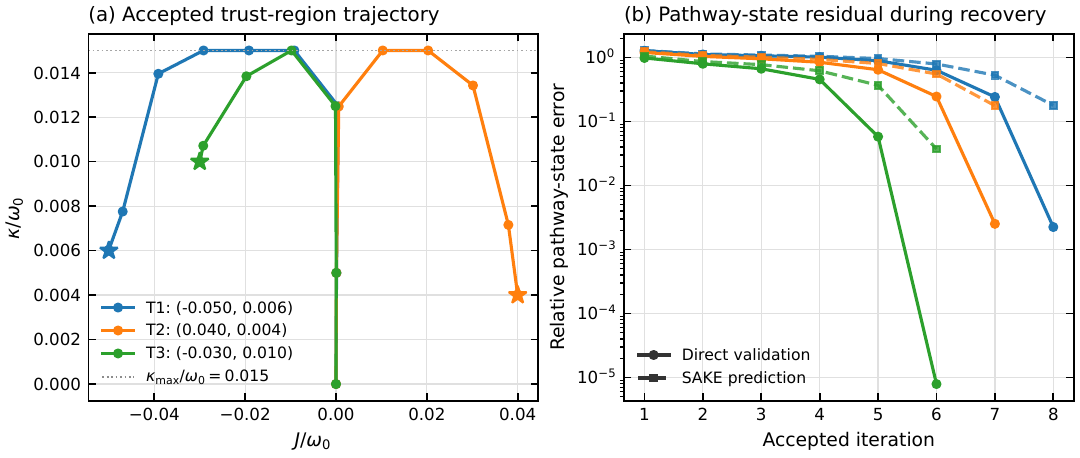}
  \caption{\textbf{Iterative inverse recovery with local SAKE charts.}
  (a) Accepted trust-region trajectories in the dimensionless control plane
  $(J/\omega_0,\kappa/\omega_0)$.  Stars mark the synthetic targets and the
  dotted line denotes the imposed dissipative-coupling bound.
  (b) Relative pathway-state residual versus accepted iteration.  Solid lines
  are direct-inversion validations and dashed lines are the corresponding
  third-order SAKE predictions.  All calculations begin at the uncoupled
  reference.}
  \label{fig:sake_inverse_recovery}
\end{figure*}

Figure~\ref{fig:sake_inverse_recovery} shows three recovery calculations, all
initialized at $(\lambda_J,\lambda_\kappa)=(0,0)$.  The target coordinates were
$( -0.050,0.006)$, $(0.040,0.004)$, and $(-0.030,0.010)$, corresponding to
$(J,\kappa)=(-100,12)$, $(80,8)$, and $(-60,20)$~meV, respectively.  Solid
curves in Fig.~\ref{fig:sake_inverse_recovery}(b) show the residual from direct
validation, whereas dashed curves show the residual predicted by the local
SAKE expansion.

\begin{figure*}[!t]
  \centering
  \includegraphics[width=0.98\textwidth]
  {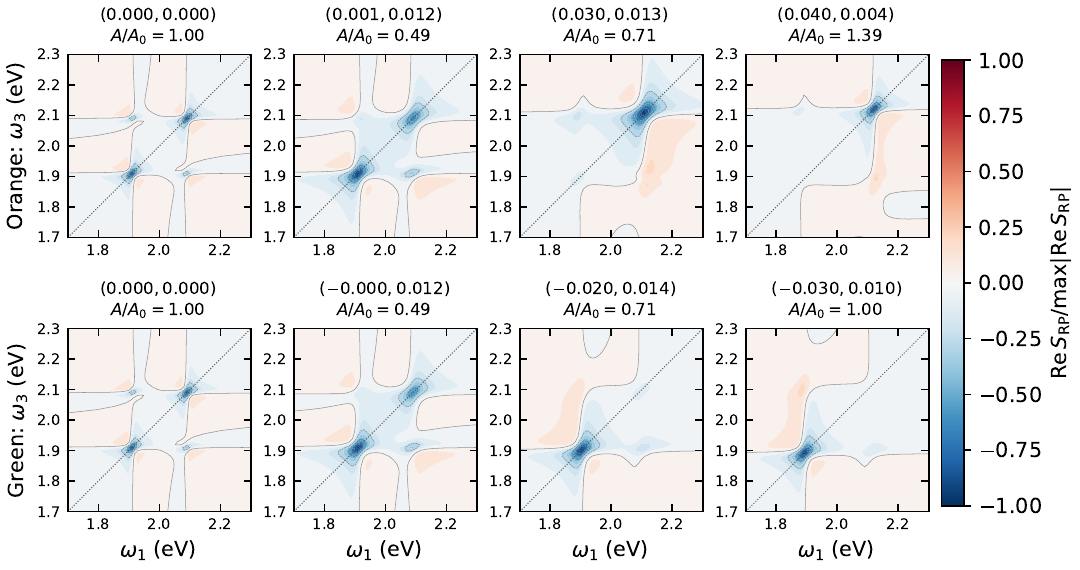}
  \caption{\textbf{Evolution of the rephasing spectrum during iterative
  recovery.} Direct rephasing spectra on a $64\times64$ frequency grid at four
  accepted chart centers along the positive-$J$ (orange, upper row) and mixed
  (green, lower row) trajectories. Coordinates above each panel are
  $(J/\omega_0,\kappa/\omega_0)$; $A/A_0$ gives the maximum absolute real
  amplitude relative to the uncoupled spectrum. The plotted signal in each
  panel is divided by its own maximum to expose changes in line shape.}
  \label{fig:inverse_spectral_evolution}
\end{figure*}

\begin{figure*}[!t]
  \centering
  \includegraphics[width=0.96\textwidth]
  {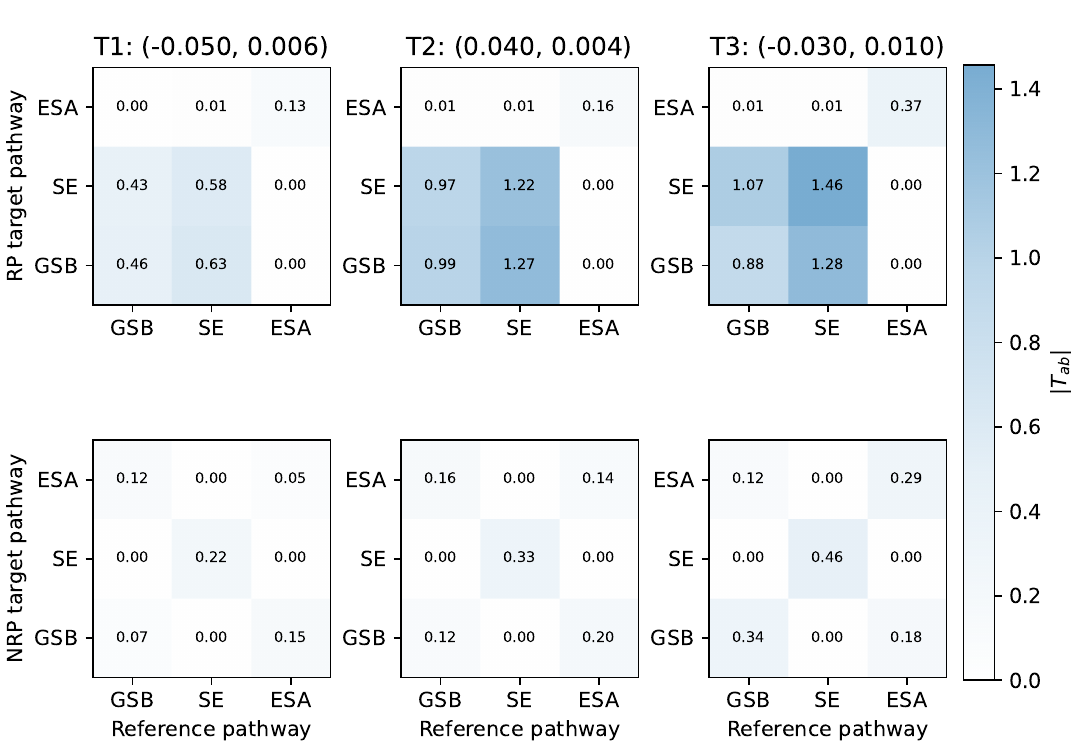}
  \caption{\textbf{Initial-to-target mixing of retained RP and NRP pathways.}
  The $3\times3$ RP (upper row) and NRP (lower row) blocks of the global
  transport matrix for the three synthetic targets. Color and the numerical
  annotation in each cell give the magnitude $|T_{ab}|$. Rows identify the
  target pathway and columns the zeroth-order reference pathway; GSB, SE, and
  ESA denote ground-state bleach, stimulated emission, and excited-state
  absorption, respectively. A common color scale is used for both sectors and
  all targets.}
  \label{fig:inverse_pathway_mixing}
\end{figure*}

\begin{table}[t]
 \caption{Synthetic recovery of coherent and dissipative dimer controls.
 Parameter errors are Euclidean distances in the physical $(J,\kappa)$ plane.}
 \label{tab:sake_inverse_recovery}
 \begin{ruledtabular}
 \begin{tabular}{cccc}
 Target (meV) & Recovered (meV) & Error (meV) & Checks \\
 \hline
 $(-100,12)$ & $(-100.026,12.068)$ & $7.26\times10^{-2}$ & 8 \\
 $(80,8)$    & $(79.967,8.058)$     & $6.72\times10^{-2}$ & 7 \\
 $(-60,20)$  & $(-60.000,20.000)$   & $2.96\times10^{-4}$ & 6 \\
 \end{tabular}
 \end{ruledtabular}
\end{table}

All three targets are recovered from the uncoupled initial model.  The number
of direct trial calculations ranges from six to eight, rather than the much
larger number of objective evaluations performed internally on the SAKE
surrogates.  The final relative pathway-state residuals are
$2.25\times10^{-3}$, $2.52\times10^{-3}$, and $7.92\times10^{-6}$ for the
three targets listed in Table~\ref{tab:sake_inverse_recovery}.  The most
distant target, $(-100,12)$~meV, follows a pronounced
$J$--$\kappa$ correlation valley and approaches the imposed dissipative bound
before turning toward the target.  This trajectory indicates that a single
spectrum may constrain combinations of coherent and dissipative parameters
more strongly than it constrains either parameter separately.  Additional
waiting times, polarization sequences, or physically motivated priors should
therefore improve identifiability in experimental applications.

Computational timings were measured on a Mac Studio equipped with an Apple M2
Max processor (12 CPU cores: 8 performance and 4 efficiency cores) with 64~GB
of unified memory using the JAX CPU backend (Python 3.13.5, JAX 0.11.0, NumPy
2.1.3, and SciPy 1.15.3).  A direct RP+NRP nonlinear response calculation at a
prescribed parameter point required approximately 2.2~s.  Construction of a
third-order local SAKE chart required approximately 9.1~s and therefore
dominates the 10--12~s computational cost of each accepted trust-region step.
In the benchmark calculations, however, only six to eight chart constructions
were required to recover the target parameters.  Once constructed, each local
chart served as a surrogate model whose objective function could be evaluated
repeatedly at negligible additional cost during the trust-region optimization.
Consequently, the computational advantage of SAKE does not arise from
accelerating an individual forward response calculation, but from amortizing
the cost of chart construction over the many objective-function evaluations
that ordinarily require repeated direct nonlinear response calculations.
Table~\ref{tab:timings} summarizes the measured cost decomposition.  Although
the absolute timings are hardware dependent, the results consistently identify
local-chart construction as the dominant computational bottleneck.

\begin{table}[t]
\caption{Representative wall-clock timings for the SAKE trust-region
implementation on a Mac Studio (Apple M2 Max, 12 CPU cores, 64~GB unified
memory) using the JAX CPU backend.  The dominant cost is construction of a
local third-order transport chart, while repeated objective-function
evaluations within that chart are inexpensive.}
\label{tab:timings}
\begin{ruledtabular}
\begin{tabular}{lcc}
Operation & Typical cost & Frequency \\ \hline
Direct RP+NRP response & $2.17$ s & Direct evaluation \\
Chart construction & $9.1$ s & Accepted step \\
Surrogate optimization & $0.044$ s & Per step (all calls) \\
Accepted-step total & $11.5$ s & 6--8 per recovery \\
Direct validation & $2.2$ s & Accepted step
\end{tabular}
\end{ruledtabular}
\end{table}

At the recovered controls the global transport matrix is obtained by directly
projecting the final pathway states onto the original zeroth-order pathway
basis.  Its diagonal elements quantify the survival or renormalization of the
zeroth-order pathways, while its off-diagonal elements quantify pathway mixing
generated during transport to the inferred model.  For the three targets the
off-diagonal Frobenius fractions are 0.708, 0.707, and 0.672, respectively,
showing that successful parameter recovery is accompanied by substantial
redistribution among the original Liouville-space pathways.

It is important to distinguish this global initial-to-final projection from a
naive product of the projected matrices from successive local charts.  The
finite reference pathway space is not exactly closed under transport: each
projection discards an out-of-basis residual, and multiplication of the local
projected matrices therefore need not be transitive.  In the present tests the
global projection is consequently used for the final mechanistic
interpretation, while the discrepancy between the global matrix and the
product of local matrices is retained as a closure diagnostic.  This
distinction does not affect trust-region parameter recovery, which is validated
against direct pathway states at every accepted step.

\subsection{Spectral evolution and pathway mixing}
\label{sec:inverse_spectral_evolution}

To connect the pathway-space diagnostic to a recognizable spectroscopic
observable, Fig.~\ref{fig:inverse_spectral_evolution} shows the directly
evaluated rephasing spectrum along the accepted trajectories to the positive-
$J$ and mixed targets.  Each panel is normalized to its own largest absolute
real amplitude so that changes in peak position and line shape remain visible;
the reported ratio $A/A_0$ retains the change in overall amplitude relative to
the uncoupled reference.  The orange trajectory transfers weight toward the
higher-frequency exciton, whereas the green trajectory increasingly emphasizes
the lower-frequency feature.  These contrasting redistributions arise even
though both trajectories initially move mainly along the dissipative-control
direction.
The corresponding initial-to-target RP and NRP transport blocks are shown in
Fig.~\ref{fig:inverse_pathway_mixing}.  A striking sector dependence emerges.
In the RP sector, the dominant off-diagonal elements mix the ground-state-bleach
(GSB) and stimulated-emission (SE) pathways, whereas excited-state absorption
(ESA) remains comparatively isolated.  In the NRP sector, the pattern changes:
the principal off-diagonal mixing occurs between GSB and ESA, while SE remains
unmixed and is only renormalized.  Variation of the coherent and dissipative
couplings therefore does not simply change pathway amplitudes; it redistributes
distinct physical response mechanisms in a phase-matching-dependent manner.
This sector-selective mixing is largely hidden after the pathways are summed
into the observable spectrum but is exposed directly by the SAKE transport
matrix.

\section{Discussion and Outlook}
\label{sec:V}

The central result of this work is a computational reformulation of nonlinear
response as local Liouvillian transport across parameter space.
Conventional response theory treats each point in parameter space as an
independent calculation: the Liouville equation is solved repeatedly
for neighboring Hamiltonians or Liouvillians, and the resulting
spectra are compared only after the fact. In contrast, the present
framework establishes an explicit relationship between neighboring models
through the transport operator
$\mathsf T_\gamma$. Once a reference response has been constructed,
neighboring responses are generated by transport rather than repeated
solution of the underlying dynamical equations.

This shift in perspective has both conceptual and computational
consequences. It transforms repeated evaluations of nonlinear
response into the problem of constructing a local transport operator.
The Spectral Autodiff Kernel Expansion (SAKE) provides a practical
realization of this idea by combining automatic differentiation with
the Duhamel transport expansion. Rather than deriving increasingly
complicated perturbative expressions analytically, the derivative
hierarchy is generated automatically by traversing the differentiable
computational graph defined by the response pathway itself.

A second consequence of the present formulation is that the transport
operator constitutes a physically meaningful object in its own right.
Traditional nonlinear spectroscopy emphasizes the measured spectrum as
the primary observable. The present work instead identifies the
pathway transport matrix as a more sensitive diagnostic of the
underlying dynamics. Its diagonal elements quantify the persistence and
renormalization of individual Liouville-space pathways, whereas its
off-diagonal elements directly reveal coherent and dissipative pathway
mixing induced by perturbations of the Liouvillian. In this sense, the
transport matrix provides mechanistic information that is largely
hidden once the individual pathway contributions are summed into the
measured spectrum.

The iterative recovery calculations demonstrate that SAKE is more than a forward transport method. The local transport expansion defines a differentiable surrogate model that can be embedded within standard trust-region optimization algorithms to recover unknown Hamiltonian and dissipative parameters from nonlinear spectroscopic observables. Although demonstrated here for a two-parameter excitonic dimer, the same strategy naturally extends to higher-dimensional inverse problems, where repeated full Liouville-space calculations become prohibitively expensive.

Although the present implementation has focused on a four-level
$\mathfrak{su}(2)\times\mathfrak{su}(2)$ excitonic dimer, the underlying framework is
considerably more general. The construction depends only on the
existence of a differentiable map from model parameters to the
Liouvillian generator and therefore extends naturally to higher-order
nonlinear spectroscopies, larger excitonic systems, vibronic models,
spin networks, cavity QED, and other open quantum systems. Because the
transport hierarchy is generated automatically, the complexity of the
analytical derivation no longer grows combinatorially with response
order.

The differentiable transport viewpoint also suggests several new computational
directions. Since the response is represented as a map
over parameter space, gradients with respect to Hamiltonian and
dissipative parameters become directly available for parameter
estimation, inverse spectral design, optimal quantum control, and
gradient-based optimization. Likewise, the transport operator provides
a natural reduced representation of neighboring models that may prove
useful for surrogate modeling and machine-learning approaches to
spectroscopic prediction.

More broadly, nonlinear spectroscopic pathways may be treated not merely as
algebraic terms in a response expansion, but as differentiable computational
graphs connected by local Liouvillian transport. Automatic differentiation
supplies the derivative tensors needed to traverse these graphs efficiently,
while the pathway-space representation preserves a direct physical account of
how perturbations redistribute nonlinear response.

\begin{acknowledgments}
CSA acknowledges funding from the Government of Canada (Canada Excellence Research Chair CERC-2022-00055),
the Institut Courtois, Facult\'e des arts et des sciences, Universit\'e de Montr\'eal (Chaire de recherche de direction de l'Institut Courtois), and the Natural Sciences and Engineering Research Council of Canada (NSERC Discovery Grant RGPIN-2024-05893).
ERB acknowledges funding from the National Science Foundation (CHE-2404788), the Robert A.\ Welch Foundation (E-1337), and the U.S.\ Department of Energy, Office of Science, under Award No.\ DE-SC0025706. ERB gratefully acknowledges funding from IVADO for a Visiting Professorship at the Institut Courtois, Universit\'e de Montr\'eal.
\end{acknowledgments}

\subsection*{Use of Generative Artificial Intelligence}
In compliance with institutional guidelines of the Universit\'e de Montr\'eal and the 
University of Houston, generative artificial intelligence tools were used to assist with the editing of language and stylistic refinement of parts of the manuscript and to assist in the synthesis of the literature. The SAKE-DT code was developed, tested, and validated using Codex (v5.5). 
These tools were not used to generate scientific content, perform analysis, or influence the interpretation of results. All content has been reviewed and validated by the authors, who assume full responsibility for the manuscript.

\section*{Data Availability}
The code and numerical validation data supporting this work are openly available in the \href{https://github.com/ebitnet65/Duhamel_Transport}{\texttt{Duhamel\_Transport}} repository, on the SAKE-DT autodiff release branch \texttt{codex/sake-autodiff-dimer}. The archived release is deposited on Zenodo with DOI \href{https://doi.org/10.5281/zenodo.21793438}{10.5281/zenodo.21793438}.
This archive contains the SAKE-DT implementation, including model-agnostic Liouville-space transport routines, JAX/autodiff Liouvillian derivative tools, Duhamel resolvent and propagator expansion routines, tutorial notebooks, benchmark scripts, and validation workflows for the four-level exciton-dimer model.

\appendix

\section{SAKE-DT Package Workflow}
\label{appendix:C}

The calculations reported in this work were performed using the
\texttt{SAKE-DT} development branch of the
\texttt{Duhamel\_Transport} package, available through the GitHub
repository identified in the Data Availability Statement. The package
implements the workflow illustrated schematically in
Fig.~\ref{fig:elementary_dag}, separating model specification,
Liouvillian differentiation, Duhamel transport, and spectral
reconstruction into independent modules. A typical calculation
proceeds through the following stages.

\paragraph{Stage 1: Model specification.}
First, the user defines the model by supplying Hilbert-space operators:
a Hamiltonian \(H(\mathbf{p})\), a set of Lindblad collapse operators
\(\{C_k(\mathbf{p})\}\), an initial density matrix \(\rho_0\), and the optical
perturbation operator \(\mu\).  These are wrapped in a
\texttt{LiouvilleModel} object, which converts the operator-level model into a
Liouville-space generator.  The user then specifies a \texttt{ParameterMap},
which contains the reference parameter dictionary and the subset of physical
parameters promoted to dimensionless control coordinates
\(\boldsymbol{\lambda}\).

\paragraph{Stage 2: Automatic differentiation.}
For the SAKE transport calculation, the same model is supplied as a
JAX-compatible Liouvillian map,
\[
\boldsymbol{\lambda}
\mapsto
\mathcal L(\boldsymbol{\lambda}),
\]
from which automatic differentiation evaluates
\(\mathcal{L}^{(0)}\), \(\mathcal{L}_{,\mu}\),
\(\mathcal{L}_{,\mu\nu}\), and, when requested,
\(\mathcal{L}_{,\mu\nu\kappa}\).  These derivative tensors are passed to the
Duhamel layer rather than finite-differencing complete response functions.

\paragraph{Stage 3: Duhamel transport.}
The module \texttt{Liouville\_duhamel.py} constructs derivative expansions of the
resolvent and waiting-time propagator.  These propagator derivatives are
combined with the dipole interaction sequence defining each phase-matched
pathway. The resulting reference pathway states and their derivatives are
projected onto the operational pathway basis at
$\boldsymbol{\lambda}=0$ by
\texttt{Liouville\_transport\_expansion.py},
yielding the Taylor coefficients of the pathway transport operator
\(
T_a{}^b(\boldsymbol{\lambda}),
\)
which is subsequently used to reconstruct neighboring nonlinear
responses.

\paragraph{Stage 4: Spectral reconstruction.}
Finally, the driver scripts evaluate the transported pathway signals
and reconstruct the nonlinear spectra. The dimer benchmark discussed
in Sec.~IV is implemented in
\texttt{demo\_dimer\_duhamel\_transport.py},
while
\texttt{benchmark\_dimer\_transport\_backends.py}
compares finite-difference and autodifferentiated derivative
backends. The accompanying tutorial notebooks expose the same
workflow interactively and provide templates for user-defined
Liouvillian models.

The following pseudocode illustrates the minimal package-level setup for an
autodifferentiated Duhamel transport calculation.  The user supplies the
Hilbert-space model, selects a reference parameter set and control coordinates,
and then passes the resulting Liouvillian derivative tensors to the Duhamel
transport layer.

\begin{widetext}
\begin{lstlisting}[language=Python,caption={Minimal SAKE-DT input structure.}]
# -------------------------------------------------------
# Minimal SAKE-DT workflow
# -------------------------------------------------------
# Stage 1: Define model
model = LiouvilleModel.from_hamiltonian(
    dimension=dim,
    hamiltonian=H,
    collapse_ops=c_ops,
    initial_state=rho0,
    transition_dipole=mu,
)

# Stage 2: Define transport coordinates
parameter_map = ParameterMap(
    reference_params=p_ref,
    controls=(
        ControlSpec("J_12", p_ref["J_12"], delta_J),
        ControlSpec("k_12", p_ref["k_12"], delta_k),
    ),
)

# Stage 3: Autodifferentiate Liouvillian
L_derivs = jax_liouvillian_derivatives(
    jax_liouvillian,
    parameter_map,
    liouville_dimension=model.liouville_dimension,
    max_order=3,
)

# Stage 4: Construct Duhamel transport
expansion = duhamel_transport_expansion_2d(
    model,
    parameter_map,
    omega1,
    omega3,
    pathways,
    liouvillian_derivatives=L_derivs,
    max_order=3,
)
\end{lstlisting}
\end{widetext}

\bibliography{Refs-Local-clean,auto-diff-refs}

\end{document}